\documentclass[11pt,conference,compsoc,a4paper,onecolumn,twoside]{IEEEtran}

\usepackage{tracefnt}
\usepackage{amsmath}
\usepackage{graphicx}
\usepackage{afterpage}
\usepackage{pifont}
\usepackage{geometry}
\usepackage{vmargin}
\usepackage[T1]{fontenc}
\usepackage[latin1]{inputenc}
\usepackage{listings}
\usepackage{subfigure}
\usepackage{url}
\usepackage{picins}
\usepackage{multirow}

\setmargins{3.0cm}{3.0cm}       
           {15.0cm}{23.7cm}     
           {0pt}{0pt}         
           {0pt}{30pt}          

\begin{document}
\bstctlcite{myctlfullname}

%
 \renewcommand{\thesection}{\arabic{section}}
 \renewcommand{\thesectiondis}{\thesection.}
 \renewcommand{\thesubsectiondis}{\thesectiondis\arabic{subsection}.}
 \renewcommand{\thesubsection}{\thesection.\arabic{subsection}}
 \renewcommand{\thesubsectiondis}{\thesubsection.}
 \renewcommand{\thesubsubsectiondis}{\thesubsectiondis\arabic{subsubsection}.}
 \renewcommand{\thesubsubsection}{\thesection.\thesubsection.\arabic{subsubsection}}

\title{Inter-organizational fault management: Functional and organizational core aspects of management architectures}
 
\author{
\IEEEauthorblockN{Patricia Marcu, Wolfgang Hommel}
\IEEEauthorblockA{Leibniz Supercomputing Centre\\
Boltzmannstr. 1, 85748 Garching, Germany\\
\{marcu, hommel\}@lrz.de}

%
}

\maketitle

\begin{abstract}

\textit{Outsourcing -- successful, and sometimes painful -- has become one of the hottest
topics in IT service management discussions over the past decade. IT services are outsourced to 
external service provider in order to reduce the effort required for and
overhead of delivering these services within the own organization.  
More recently also IT services providers themselves started to either outsource service parts 
or to deliver those services in a non-hierarchical cooperation with other providers. 
Splitting a service into several service parts is a non-trivial task as they have to be 
implemented, operated, and maintained by different providers. One key aspect of such 
inter-organizational cooperation is fault management, because it is crucial to locate and 
solve problems, which reduce the quality of service, quickly and reliably. 
In this article we present the results of a thorough use case based requirements 
analysis for an architecture for inter-organizational fault management (ioFMA). 
Furthermore, a concept of the organizational respective functional model of the ioFMA
is given.
}

\end{abstract}


\section{Introduction}\label{sec:1motiv}

Providing IT services in an inter-organizational manner is a complex and often error-prone task. Managing IT services is often characterized by applying the classic FCAPS partitioning: fault, configuration, accounting, performance, and security management. In this article, we focus on the technical functionality as well as the organizational aspects of fault management in the context of inter-organizationally operated IT services. 
Our work is primarily motivated by the interaction of the following three challenges:

\paragraph*{\textit{The \textbf{Outsourcing} problem}}: 
One characteristic of the last decade is that many organizations have outsourced their IT services to external parties, either entirely (e.g. email, file storage, and web servers) or just partially. Consequently, many processes and workflows have been transferred to and restructured by these external service providers. Outsourcing is performed in order to reduce the organization's IT costs, but also to facilitate good technical support. Related to these goals, ITIL v3 (see \cite{itil07d}) describes the migration from the \textit{value-chain-model} -- also known as hierarchical service delivery model -- to the \textit{value-network-model}, which contains horizontal (non-hierarchical) relationships between the involved providers. Within this scope, different sourcing strategies are defined.

\paragraph*{\textit{The problem of \textbf{heterogeneity and autonomy} in multi-domain environments}}:
From an organizational point of view, IT service providers collaborate with each other in very diverse ways. This makes it difficult to specify a single, universal methodology for effective and efficient inter-organizational fault management (\emph{ioFM})
The common denominator of the organizational models found in practice is the \textit{heterogeneity} and \textit{autonomy}, especially concerning the deployed IT systems and management tools. We therefore have to face the challenge of specifying fault management concepts that are to be deployed in cross-organizational or multi-domain environments and deal with these characteristics.

\paragraph*{\textit{The \textbf{service delivery diversity}}}:
Regarding the service delivery process as a productive process, a big difference between the real organizations concerning  process control, communication, and many other IT service management (ITSM) aspects can be observed. Therefore very useful reference processes exist.
Reference processes for fault management have been described, for example, in \cite{itil07d} and \cite{etom} for hierarchical service delivery, and in \cite{Hamm09} for heterarchical (i.e. non-hierarchical) service delivery. Based on these reference processes, other related work, and real-world scenarios, we have extracted the requirements for an ioFM architecture as presented in this paper.

\begin{figure}[] 
	\centerline{\includegraphics[width=0.75\columnwidth]{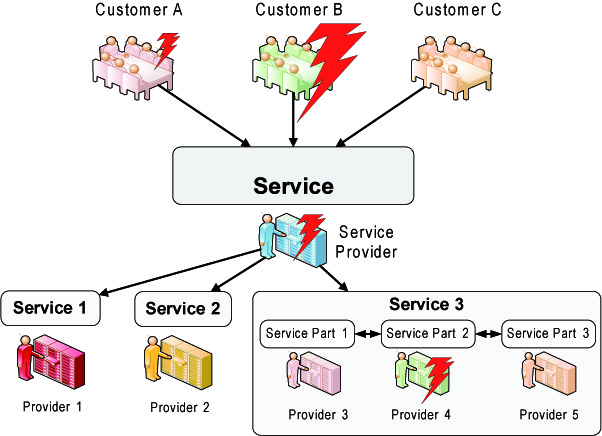}}
	\caption{Propagation of faults in inter-organizational environments}
	\label{fig:complex} 
\end{figure}

The above stated problems are those characteristics of inter-organizational IT environments that are most relevant for ioFM; a simple example is given in Figure \ref{fig:complex}: A provider delivers its services to customers A, B, and C in different ways: It outsourced three of the services (labeled Service 1, 2, and 3 respectively) to other service providers. Up to this point we deal with a vertical service chain, which represents the classical type of \textit{hierarchical service delivery}. Both services 1 and 2 are delivered by only one service provider (providers 1 and 2 are the subcontractors of the service provider). Opposed to these two services, Service 3 is provided by multiple cooperating service providers (Provider 3, 4, and 5). Each of these three providers is required to deliver its part of the service, but none of them has a superior role; instead, they are on a par with each other: These service providers coexist on the same ,,service layer'' regarding the service functionality. They deliver ,,service parts'' (as discussed in \cite{hmy08}) (Service Part 1, 2, and 3 respectively) which together lead to the delivery of a single horizontal service. These service parts are concatenated within the same service layer, so the horizontal service chain represents a \textit{heterarchical service delivery}. 


It is usual that each real world organization aligns itself on its own requirements, workflows, and processes. It also uses different IT infrastructures, systems, and tools. As a consequence, each organization we deal with needs to be analyzed first, and typically there is a lack of tool interoperability whenever multiple service providers are about to be coupled in order to jointly provide an IT service. In this context, management tool support is of utmost importance, because the complexity of the IT infrastructure as well as of each service increases with the number of involved providers.

Taking into account the above stated challenges, the scenario described here is clearly a heterogeneous one. Following issue is important here: 
A fault, e.g., within the Provider 4's domain, will -- independent of its root cause -- make the whole Service 3 fail because of this issue within Service Part 2. This fault will be propagated to the Service Provider, and thus the customers will face a quality-degraded or unavailable service. This fault can have more or less follow-ups depending on the service customization for each individual customer. Nevertheless, in such inter-organizational scenarios it is very difficult to precisely locate such a fault, to correlate it with other unsolved faults, and to track and steer the progress of the handling and correction.

For a single IT service provider's infrastructure already several approaches and best-practices concerning fault management exist. But regarding ioFM there is a lack of both research and best practices.
Our work faces the additional practical challenge that IT service providers from different countries are involved, which in turn increases both the technical complexity as well as the organizational and legal constraints, resulting in even more complex delivery processes. 

Regarding outsourcing as well as multi-domain IT service delivery from a process-oriented point of view, a well defined and proper ioFM is needed on the \emph{system layer}. In order to meet this demand, our work focuses on an ioFM Architecture (ioFMA). This article presents our methodology and the results of our ioFMA requirements analysis. It is structured as follows: In Section~\ref{sec:2relwork} we sum up the related work that has influenced our methodology and ioFMA design. In Section~\ref{sec:3design}, we present details about our design rationale and the MDA-based approach that has been taken. Section~\ref{sec:4scen} outlines the inter-organizational scenarios we have analyzed. Section~\ref{sec:4req} specifies the roles and actors relevant to ioFM on which the organizational model bases and on this basis we then present the identified use cases and the derived requirements. In Section~\ref{sec:6funcModel} we are giving an overview on the functional model of the ioFMA. A summary and an outlook to our future work concludes this paper in Section~\ref{sec:5further}.

\section{Related Work}\label{sec:2relwork}

\subsection{Management architectures and their submodels}

In Hegering et al. \cite{han99} the building blocks of \textit{management architectures (MA)} are described. The primary goal of each management architecture is to establish an \emph{integrated} management approach by providing a valid system management framework instead of using several management tools independently of each other. The MA is composed of four complementary submodels: the \textit{information model (IM)}, the \textit{organizational model (OM)}, the \textit{communication model (CM)}, and the \textit{functional model (FM)}. The \textit{IM} represents the description and modeling of the managed objects (management-relevant information to be exchanged). The \textit{OM} describes the roles as well as the responsibilities and specifies the communication patterns within the \textit{MA}. The \textit{CM} specifies the communication procedures for the exchange of management information. The FM splits the management task into several components and provides dedicated management functionalities: \textit{fault management, configuration management, accounting management, performance management, and security management (also known as FCAPS)}.

The MA concept along with its submodels is very valuable for this work, because it the base for holistic integrated network management. Thus our work will be aligned to the four submodels of such a MA. They have to be extended to take inter-organizational conditions into account, which have not been considered by previous MA variations yet. Also the functional area of fault management (FM) will be taken into account and refined to additional ioFM functionalities that are tailored for inter-organizational environments.  

\subsection{IT Service Management}

ITSM frameworks, such as ITIL v3 \cite{itil07d}, ISO/IEC 20000 \cite{iso20000}, and eTOM \cite{etom} have been established to design management processes that follow the continual improvement strategy of Deming's plan-do-check-act life cycle. These ITSM frameworks have been used primarily for process definition in hierarchical service delivery scenarios. For non-hierarchical service delivery, a new concept has been developed in \cite{Hamm09}. 

These approaches give guidelines for the inter-organizational service delivery \emph{processes} as a whole. Nevertheless, on the (technical) \emph{system layer} there is no underlying concept for inter-organizational service delivery defined yet. Our work focuses on refining the given reference processes and designing an integrated system-level MA.

\subsection{Service Composition}

As we take into account services delivered in an inter-organizational environment, the concept of service composition is a key enabler for our research. In \cite{Dreo02}, Dreo distinguishes between two types of supply chains: vertical and horizontal. By vertical the well known hierarchical service delivery is meant. The horizontal supply chain addresses the issue of peering. 
Despite the partially overlapping scope between these results and our work, the non-hierarchical service delivery taken in account by our research does not only cover peering. The underlying necessity has also been postulated by Hedlund \cite{Hed05}, whose work uses the term \emph{heterarchy} for the non-hierarchical organizational forms, which we also address.

In their work \cite{vpmb07} on service composition applied to network management, Vianna et al. show that service composition can indeed be realized by using traditional management technologies. The application of technologies created to support service composition will bring important advantages to the network management discipline. However, they consider only services based on a hierarchical chain of compositions. 

Klie et. al analyze the automatic web service composition as a possibility to further automate network management in \cite{kgf07}. They compare several web service composition technologies in order to describe an approach using a composition engine for network management. This automatic web service composition can be used to simplify complex network management tasks. It also enables the automatic composition for covering large parts of several network management tasks; this approach is valuable as a guideline for the implementation of the ioFMA.

\subsection{Fault Management related Tasks}

In \cite{mdkp09} a framework for problem determination is proposed.  It is based on the monitoring of event streams that are generated by the different components of an IT service. A generic representation of a problem through spatial-temporal patterns is given. Additionally, efficient algorithms are described in order to sustain building blocks for a hierarchical heuristic for detecting generic patterns. Even though some of these concepts are distantly related to our approach, their work is merely based on hierarchical service structures. Also in \cite{GPM08a} the automation of the incident management is proposed.

In our former work \cite{MSGL09} we specified a methodology for handling faults in non-hierarchical service delivery environments, which we called Service Provider Coalitions. This approach's goal was the correlation of fault reports generated by different incident ticketing systems in multi-enterprise environments. We now propose to realize the fault management on a higher level of abstraction.

\section{Design rationale}\label{sec:3design}

This section describes the methodology used in designing the architecture, several of the taken design decisions, and the consequences for the ioFMA. 

\subsection{Model Driven Architecture}\label{subsec:mda}
\begin{figure}
	\centering
  	\includegraphics[width=0.75\columnwidth]{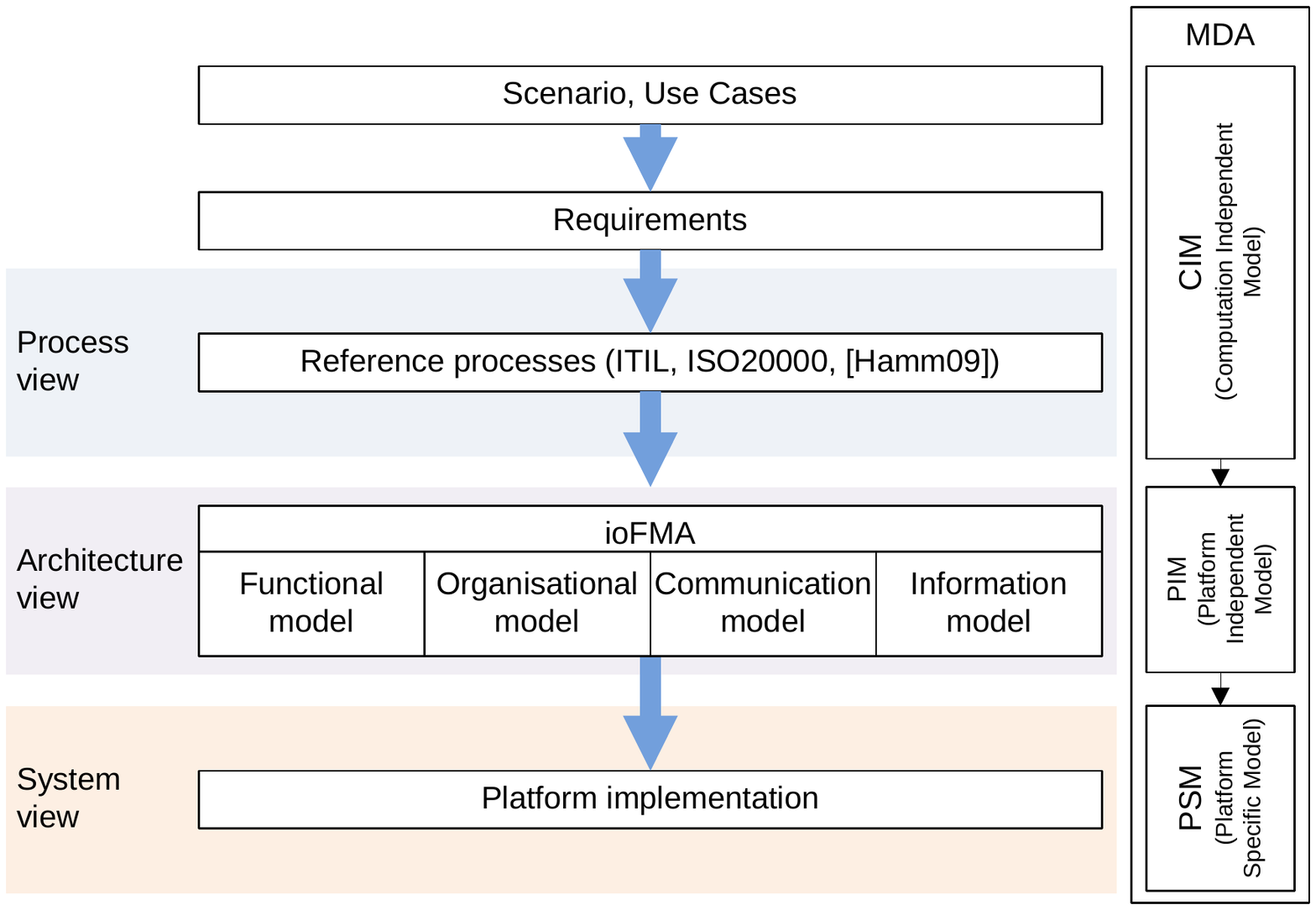}
	\caption{Our methodology, which is based on the MDA approach}
	\label{fig:mdaallg} 
 \end{figure}

Our design of the management architecture follows the \textit{Model Driven Architecture (MDA)} \cite{OMG03} approach. Its iteratively refining character is outlined in Figure \ref{fig:mdaallg}. 
MDA contains three models: 

\begin{enumerate}

\item The \textit{computation independent model (CIM)} provides a general view on the system, as well as on the environment in which this system will be deployed.

\item The \textit{platform independent model (PIM)} provides a view on the system independently of the platform that it will be deployed on. Consequently, this model is still generic and can be applied to several platforms of similar type.

\item The \textit{platform specific model (PSM)} takes the specification from the PIM and describes its application to a specific platform.

\end{enumerate}

As a result, the three models build upon each other and descend from a higher level of abstraction (CIM) to a lower one (PSM).
The design of our ioFMA is done in analogy to MDA. In our design process, the requirements elicitation and its model design correspond to MDA's CIM view.  

The scenarios' description (one hierarchical and one heterarchical scenario) and their generalization are part of the requirement analysis. From the resulting general scenario we derive use cases and several implicit requirements on the ioFMA.

A three-tier procedure for the model design is used:

\begin{enumerate} 

\item The \textit{process view} corresponds to CIM and contains reference processes regarding Incident Management (for hierarchy we used \cite{itil07d} and for heterarchy we used \cite{Hamm09}).

\item The \textit{architecture view} corresponds to PIM and contains the ioFMA as well as its sub models, which correspond to the described processes in the upper layer. 

\item The \textit{system view}, which is representing the PSM in our approach, contains the implementation of the overlaid architecture on any specific platform on the system layer.

\end{enumerate}

Furthermore, the design methodology of our ioFMA is split into two parts: the requirements analysis and the model design.


\subsection{Methodology of requirements analysis}

In order to elicit ioFMA requirements, we have analyzed two real world scenarios: The IntegraTUM scenario as an representative example of a hierarchical inter-organizational service delivery, and the G\'EANT scenario representing the heterarchical service delivery in inter-organizational environments. Based on these practical scenarios, we derived a more abstract generic scenario and its use cases. The textual description of the use cases has been performed with a focus on management architectures (cf. section \ref{sec:2relwork}) and their sub models. Functional and non-functional requirements have then been derived from these use cases.

\subsection{Methodology of model design}

Based on the requirements and on the reference process for incident management (cf. section \ref{sec:2relwork}), the sub models of our ioFMA are specified in the following order: 

\begin{enumerate}

\item The \textit{functional model}, which has to underline the most important functionalities concerning fault management, comes first.

\item The \textit{organizational model} follows and reveals the roles and responsibilities in inter-organizational environments that are required in order to conduct efficient ioFM.

\item The \textit{communication model} then delivers the required information communication exchange measures and procedures.

\item The \textit{information model} finally specifies the data format for the ioFM information exchange and processing.

\end{enumerate}

In the next step, the \textit{ioFMA} will be transformed to a PIM; then it will be instantiated for hierarchical, heterarchical, and mixed forms of service delivery. All of them will be mapped onto PSMs. 
In the next section, we present details about the first step in this methodology, i.e. the requirements analysis.

\section{Scenarios for inter-organizational fault management}\label{sec:4scen}

In order to design and implement an ioFMA, we have chosen the following two scenarios, one for each inter-organizational service delivery model: hierarchy (IntegraTUM) and heterarchy (G\'EANT).

\subsection{IntegraTUM}
In the IntegraTUM project \cite{IntegraTUM}, which has been funded by the German Research Foundation (DFG) and initiated by the Technische Universit\"at M\"unchen (TUM), several university IT services, which were previously operated by the various TUM institutions (e.g. library, administration, and faculties) themselves, have been reorganized and recentralized at the Leibniz Supercomputing Center (LRZ).

TUM's staff and students are automatically granted access to all relevant services, such as the university web portal, learning management system, and computer labs based on an identity management process that is coupled with the student enrolment process and the human resources (HR) management software. Thus, TUM is LRZ's customer and the scenario fulfills the criteria of the hierarchical inter-organizational service delivery model as outlined above. 
A fault management process has been established between the both organizations in this hierarchy and is described in detail in \cite{hokn10}.

\subsection{G\'EANT}
The End-to-End (E2E) Link service in the G\'EANT2 multi-national network \cite{Geant10} is an example of services delivered by a heterarchical service provider organization.
 
Co-funded by the European Commission as well as Europe's national research and education networks (NRENs), and managed by DANTE, the G\'EANT network connects 34 countries via 30 NRENs. On the technical layer, multiple 10Gbps wavelengths are used to set up dedicated E2E links. One representative customer is the Large Hadron Collider (LHC) project at CERN in Switzerland. It is expected that its recently started experiments will produce 15 petabytes of scientific data each year. In order to meet the bandwidth and quality of service requirements of large-scale research projects, dedicated optical E2E Links must be set up. These links span multiple countries and allow the unrestricted utilization of the physically possible bandwidth. 

E2E Links connect organizations located in different countries and cross the networks of different providers. When providing the E2E Link services, each provider (member of the service provider coalition) has to collaborate w.r.t. setup, maintenance, and management with the other providers. Major challenges in the realization of these services are the heterogeneity concerning the technical implementations, the used software tools, various people related issues, and many more.
In \cite{Hamm09}, Hamm introduced a reference incident management process for E2E Links.


\section{Use cases and requirements elicitation}\label{sec:4req}

Both of the scenarios outlined above provide plenty of use cases for the elicitation of ioFMA requirements, although fault management obviously is only one of a lot of aspects that need to be addressed in such complex service provider constellations. One of the characteristics common to both scenarios is that the service providers, which are involved in the delivery process, are communicating and cooperating with each other in a kind of \textit{,,provider network"}. To better address such specifics, we first define the roles for ioFM in the next section. They have been generalized based on the roles and responsibilities we found in the real world scenarios.

\subsection{Defining roles for inter-organizational fault management}\label{subsec:roles}

One of the most important roles in ioFM is the \textit{user}. This is the role that typically initiates the fault management process by means of fault notifications that are stored in trouble ticket systems (TTS). In inter-organizational environments this role can be assigned to a service provider that is using a certain service as a \textit{user}, e.g., due to outsourcing. 

\textit{Service Provider (SP)} is the role that is responsible for the delivery of a service and for the fulfillment of the Service Level Agreements (SLAs) agreed with its \textit{users}. These \textit{SP}s are also essential to the ioFM as they constitute the \textit{provider network} and deliver IT services in a cooperative manner.

Within the different service provider domains, there is always a role that is responsible for the local fault management. We called this role the \textit{Domain Fault Manager (DFM)}. The \textit{DFM} does not only communicate within its domain, but also with the \textit{DFM}s of other domains.

On the local level also a \textit{Domain Fault Operator (DFO)} is required in order to isolate, correct, and log a fault within her own domain. Even though these both are intra-organizational roles, the \textit{DFO} has a purely operational role, whereas the \textit{DFM} primarily has coordinating responsibilities.

In ioFM, the so-called \textit{Global Fault Coordination Manager (GFCM)} has the overall coordination role: It addresses all the domains that are involved in the service delivery process. The GFCM's main tasks include: monitoring of confirmed and potential faults, forwarding of fault-related information between the different domains, and facilitating inter-domain communication. In the hierarchical case the role of the \textit{GFCM} is identical to \textit{DFM} for obvious reasons. However, in a heterarchy, the role of \textit{GFCM} will be assigned temporarily to each of the domains in an on-demand manner.

Last but not least the \textit{Domain Monitoring System (DMS)} is responsible within a domain for system and component monitoring. This role announces fault notifications or alarms about malfunctions of the system. Using these roles the use cases are described in the following section.

The important roles defined here are the base for the \textit{organizational model} of the ioFMA.

\subsection{Identifying use cases}\label{subsec:usecase}

Above we describe and analyze the two real-world scenarios in order to elicit use cases needed for the requirements analysis.
Therefore we have identified the following different classes of use cases: \textit{fault localization}, \textit{fault resolution progress management}, \textit{monitoring}, \textit{reporting}, and \textit{handling false-positives}. These also represent the core functionalities that an ioFMA should offer.


\begin{figure}[t]
\centering
\subfigure[Use cases for fault localization]
 {\includegraphics[width=0.49\textwidth]{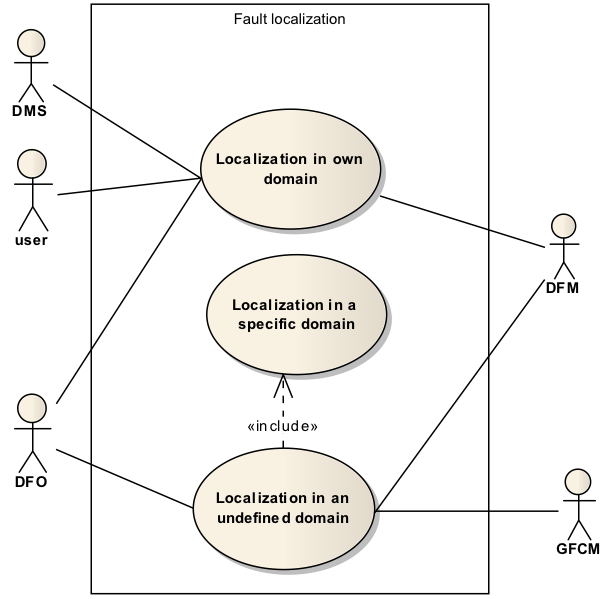}\label{fig:ucfaultloc}}
\subfigure[Use cases for fault monitoring]
 {\includegraphics[width=0.49\textwidth]{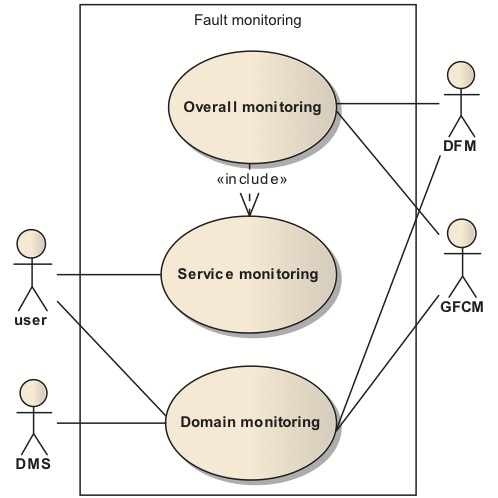}\label{fig:ucmonitor}}
\caption{Use cases for fault localization and monitoring}
\end{figure}

\subsubsection{Fault Localization}
The main functionality of the ioFMA has to be the precise localization of faults. Depending on the place where the fault will be localized, there can be multiple variations as shown in Figure \ref{fig:ucfaultloc}: 
The \textbf{\textit{fault localization within one's own domain (L01)}} is initiated by the \textit{user}, or by the \textit{DMS} respectively, and will be localized by the \textit{DFM} if a known fault occurs; otherwise, i.e. if it is an unknown fault, it will be the \textit{DFM}'s task with the support of the \textit{DFO}. 

If the fault cannot be isolated within this domain, the issue will be forwarded to another domain. The \textbf{\textit{fault localization in an undefined domain (L02)}} will therefore be initiated. The \textit{DFM} is reporting the fault to the \textit{GFCM}, which will forward it to all \textit{DFM}s involved in the service delivery. In collaboration with the \textit{DFO}s, the fault will -- in the best case -- be found in one of the domains and back reported to the \textit{GFCM}. However, in the case that the fault cannot be isolated in this way, an escalation procedure has to be initiated. 
A derivate of this use case is \textbf{\textit{fault localization within a specific domain (L03)}}; here, the \textit{GFCM} has to forward the fault only to a certain (known) domain and not to all involved partners.

\subsubsection{Fault Resolution Progress Management}
A status display informs about the \textit{progress of the fault resolution} or the \textit{progress of the maintenance work}. The \textbf{\textit{progress of the fault resolution (P01)}} is initiated by the \textit{DFM} that wants to know the progress of the fault resolution within his own or any other involved domain. It can also be initiated by the \textit{GFCM} in order to get an overview of the whole inter-organizational network with respect to the fault resolution process instances. Consequently, the \textit{DFM} and/or \textit{GFCM} query the \textit{DFM}s regarding the progress of the fault resolution in their respective domains. The \textit{DFM}s will retrieve this information from their \textit{DFO}s and give feedback to the \textit{DFM} or \textit{GFCM} from which the query originates. For the \textbf{\textit{progress of the maintenance work P02}} the same steps will be run through, but with a different scope. The case when a user wishes to be informed about the status of the fault resolution and/or maintenance is a secondary scenario within this use case, which results in a query forwarded by the \textit{DFM} or \textit{GFCM}.


\subsubsection{Monitoring}
In both the hierarchical and the heterarchical case, monitoring is a very important feature that the ioFMA should have. By means of continuous monitoring, faster fault localization is enabled. We distinguish between \textit{domain monitoring}, \textit{overall monitoring}, and \textit{service monitoring} (see figure \ref{fig:ucmonitor}). The \textbf{\textit{domain monitoring (M01)}} is responsible for the fault monitoring within a domain. It can be initiated by the \textit{user}, \textit{DFM}, or \textit{GFCM}. They will be querying the DFM of a certain domain about the general status of the faults within this domain. The result will be retrieved from the \textit{DMS}, which is always updated concerning the alarms and fault notifications. One exception that needs to be dealt with is when the \textit{user} or \textit{DFM} does not have the necessary access rights to fetch monitoring information about another domain. \textbf{\textit{Overall monitoring (M02)}} is responsible for the monitoring of the whole provider network. It can be initiated by the \textit{GFCM} or by any other \textit{DFM} that has sufficient access rights. This results in querying the entire domain \textit{DFM}s about their monitoring status. If all of the domains are replying with a valid status, then the overall monitoring is enabled; otherwise only a partial monitoring of the provider network can be established. As many providers (but not all of those within the provider network) are involved in the delivery of a certain service, the \textbf{\textit{service monitoring (M03)}} is denoting that only these involved domains will be monitored. This is a special case of the former one, as it monitors only a well-defined subset of the provider network.

\begin{figure}[t]
\centering
\subfigure[Use cases for fault resolution progress management]
 {\includegraphics[width=0.49\textwidth]{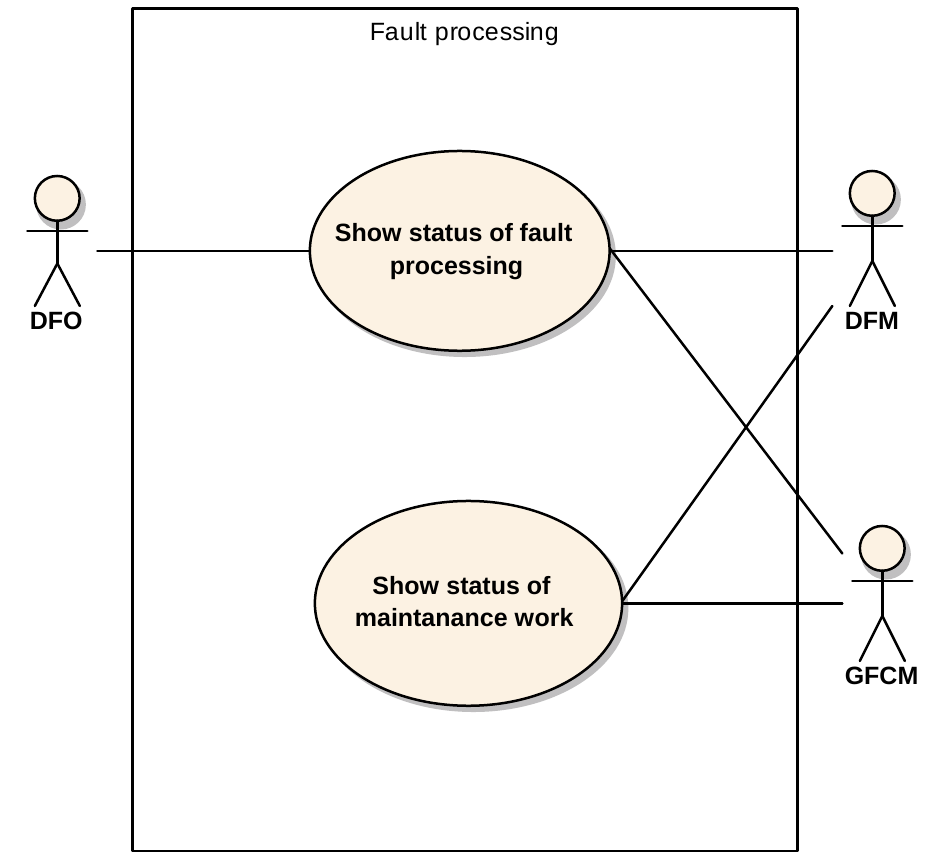}\label{fig:ucresproc}}
\subfigure[Use cases for false positives]
 {\includegraphics[width=0.49\textwidth]{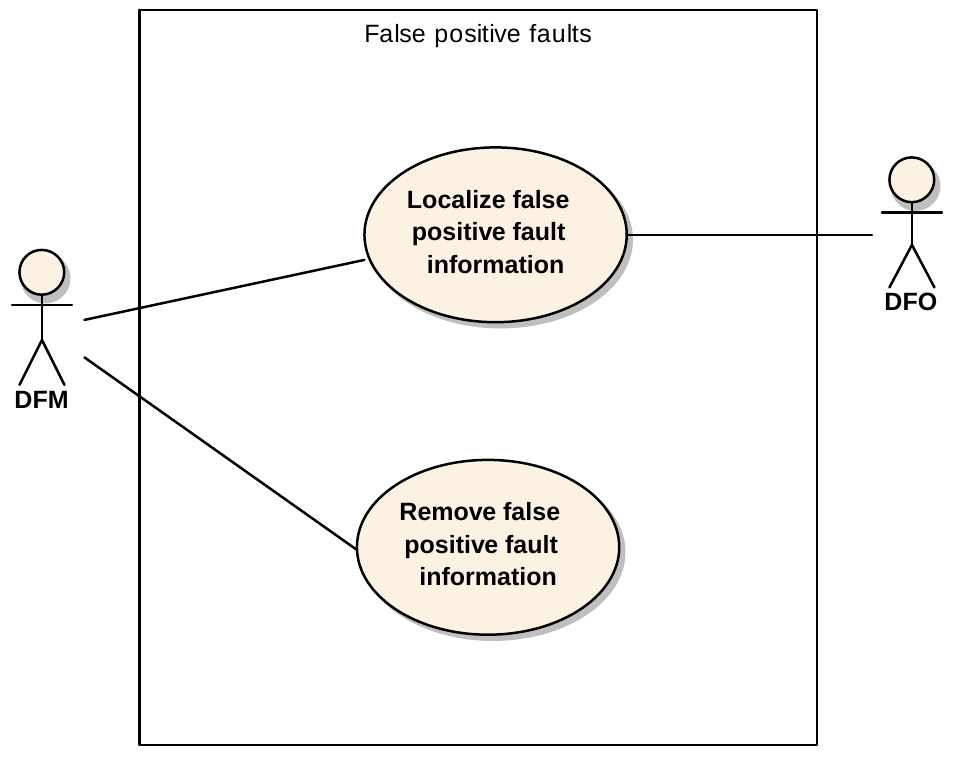}\label{fig:ucfalsepos}}
\caption{Use cases for fault resolution progress management and false positives}
\end{figure}

\subsubsection{Reporting}

Reports are supporting different processes, such as fault management. They give an \textit{overview of actual measurements, metrics, accounting data}, \textit{Quality of Service (QoS) parameters}, but also information based on historical data, e.g. in order to facilitate a \textit{trend analysis}. First the realization of\textbf{ \textit{statistical plots and accounting data reports (R01)}} will be specified. This is usually initiated by the \textit{GFCM}, which is about to retrieve all this data from all \textit{DFM}s in the provider network. In the best case all the domains send the requested information so that a report and statistical plots from all the involved domains can be conducted. In the case that some domains do not respond to the information request, incomplete statistical plots or/and accounting data will be shown. The \textbf{\textit{QoS parameter (R02)}} will be retrieved in order to check the fulfillment of the agreed SLAs and to evaluate the follow-up of different faults that have occurred in the past. Based on historical information, \textbf{\textit{trend analysis (R03)}} can be done by predicting the liability of the system to some specific faults with various follow-ups according to various statistical models. Potential future faults could therefore be resolved or by-passed before they really occur.

\subsubsection{False-positives}
In order to be assured that information concerning faults is valid, false positives (i.e. wrongly announced faults) have to be identified and removed. This use case is very important as in many cases the search for non-existing faults impedes the normal functionality of an IT service. The \textbf{\textit{localization of false positives (F01)}} is initiated by the \textit{GFCM} or by one of the \textit{DFM}s. In the case that a potential false fault notification is given that cannot be mapped onto the behavior of the system, the \textit{GFCM} or \textit{DFM} is querying the responsible \textit{DFM} about this issue. The \textit{DFM} has to consult the \textit{DFO} and figure out whether this fault really is a false positive. The result will be reported back to the GFCM. The \textbf{\textit{removal of false positives (F02)}} requires that it has reliably been identified as such first. Thus, the \textit{DFO} identifies the non-existing fault and removes the false positive (manually or tool-supported) from the monitoring system. This action is then reported to the \textit{DFM}.

\begin{table}[t]
\begin{center}
\begin{tabular}{|l|c|c|c|} \hline

& \multicolumn{3}{c|}{{\textsf{Scenario}}} \\ \cline{2-4}

\textsf{Use case} & \textsf{IntegraTUM}&\textsf{G\'EANT} &\textsf{Generalized}\\
 & \textsf{(hierarchy)}&\textsf{heterarchy} &\textsf{(mixed)}\\
\hline
\hline

\textsf{L01:}& \ding{51} & \ding{51}&\ding{51}\\ 
\textsf{L02:}&			 & \ding{51}& \ding{51}\\ 
\textsf{L03:}&	\ding{51}& &\ding{51} \\ 
\textsf{P01:}& \ding{51}& \ding{51}& \ding{51}\\ 
\textsf{P02:}&\ding{51} & \ding{51}&\ding{51}\\ 
\textsf{M01:}&\ding{51} & \ding{51}&\ding{51}\\  
\textsf{M02:}&  		& \ding{51}&\ding{51}\\ 
\textsf{M03:}&\ding{51} &\ding{51} & \ding{51}\\ 
\textsf{R01:}&\ding{51} & \ding{51}&\ding{51} \\ 
\textsf{R02:}&\ding{51}	& \ding{51}&\ding{51} \\ 
\textsf{R03:}& \ding{51}& \ding{51}& \ding{51}\\ 
\textsf{F01:}& 			&\ding{51} & \ding{51}\\ 
\textsf{F02:}& 			&\ding{51} & \ding{51} \\ 

\hline 
\end{tabular}
\caption{Coverage of the scenarios by the use cases}
\label{tab:usecase2szen}
\end{center}
\end{table}

\subsection{Deriving requirements}

Table \ref{tab:usecase2szen} summarizes the different use case occurrence as requirements for the functional model of the \textit{ioFMA}. Additional to these, the following two additional requirements have to be considered:

\begin{itemize}
\item \textbf{FM-01}: In order to increase the legibility of the fault information, a \textbf{\textit{visual presentation}} is necessary.
\item \textbf{FM-02}: Especially regarding the use cases for fault resolution progress management and in the removal of false positives the possibility to \textbf{\textit{change or remove fault data}} has to be given.
\end{itemize}

In order to support the realization of the use cases described above some requirements on the sub-models of the \textit{ioFMA} have to be fulfilled.

We identified the following requirements regarding the \emph{information model} of the \textit{ioFMA}:

\begin{itemize}
\item \textbf{IM-01}: A \textbf{\textit{common data format}} for fault information is needed in order to facilitate the inter-domain data exchange and the communication. This should consist of a set of common attributes or properties.
\item \textbf{IM-02}: Another additional or coexisting requirement to the first one is the existence of \textbf{\textit{conversion methods}} between the data format in the different domains.
\item \textbf{IM-03}: \textbf{\textit{Interface definition}} across different domains have to be defined.
\item \textbf{IM-04}: The ioFMA has to support all the \textit{\textbf{life cycle phases}} of a fault resolution process (detection, isolation, repairing/recovery, and forecast/prevention).
\item \textbf{IM-05}: Also the use of \textit{\textbf{standard metrics}} has a supporting role in the monitoring, and respectively in the reporting. An example of such a set of standard metrics is the IP Performance Metrics (IPPM) \cite{IPPM} (e.g., One Way Delay (OWD \cite{RFC2679}), IP Delay Variation (\cite{RFC3393}), Packet Loss (\cite{RFC2680}), and others).
\item \textbf{IM-06}: As the correlation/interrelation between the metrics of different domains has to be provided, a suitable \textbf{\textit{aggregation function}} has to be defined.
\end{itemize}


\begin{table}
\begin{center}
\begin{tabular}{|p{0.12\textwidth}|c|c|c|c|p{0.12\textwidth}|} \hline

\multicolumn{2}{|c|}{\multirow{2}{*}{\textsf{Requirements}}} & \multicolumn{4}{c|}{{\textsf{Phases of the fault life cycle}}} \\ \cline{3-6}
\multicolumn{2}{|c|}{}&\textsf{Detection}&\textsf{Isolation} &\textsf{Repairing}& \textsf{Forecast/\-Prevention}\\
\hline
\hline
\multirow{ 7}{*}{\textsf{for the IM}}& 

\textsf{IM-01}&&\ding{51}&&\ding{51}\\\cline{2-6}
&\textsf{IM-02}&\ding{51}&&&\ding{51}\\ \cline{2-6}
& \textsf{IM-03}&\ding{51}&\ding{51}&\ding{51}&\\ \cline{2-6}
&\textsf{IM-04}&\ding{51}&\ding{51}&\ding{51}&\ding{51}\\ \cline{2-6}
&\textsf{IM-05}&\ding{51}&&&\ding{51}\\ \cline{2-6}
&\textsf{IM-06}&\ding{51}&&&\ding{51}\\ \hline

\multirow{ 2}{*}{\textsf{for the OM}}&
\textsf{OM-01}&\ding{51}&\ding{51}&\ding{51}&\ding{51}\\ \cline{2-6}
&\textsf{OM-02}&\ding{51}&\ding{51}&\ding{51}&\ding{51}\\ \hline

\multirow{ 15}{*}{\textsf{for the FM}}& 

\textsf{FM-L01}&\ding{51}&\ding{51}&&\\ \cline{2-6}
&\textsf{FM-L02}&\ding{51}&&&\ding{51}\\\cline{2-6}
&\textsf{FM-L03}&\ding{51}&\ding{51}&&\\\cline{2-6}

&\textsf{FM-P01}&&&\ding{51}&\\\cline{2-6}
&\textsf{FM-P02}&&&\ding{51}&\\\cline{2-6}

&\textsf{FM-M01}&\ding{51}&\ding{51}&\ding{51}&\ding{51}\\\cline{2-6}
&\textsf{FM-M02}&\ding{51}&\ding{51}&\ding{51}&\ding{51}\\\cline{2-6}
&\textsf{FM-M03}&\ding{51}&\ding{51}&\ding{51}&\ding{51}\\\cline{2-6}

&\textsf{FM-R01}&\ding{51}&&&\ding{51}\\\cline{2-6}
&\textsf{FM-R02}&\ding{51}&&&\ding{51}\\\cline{2-6}
&\textsf{FM-R03}&&&&\ding{51}\\\cline{2-6}

&\textsf{FM-F01}&&\ding{51}&&\ding{51}\\\cline{2-6}
&\textsf{FM-F02}&&&\ding{51}&\ding{51}\\\cline{2-6}

&\textsf{FM-01}&\ding{51}&\ding{51}&&\ding{51}\\  \cline{2-6}
&\textsf{FM-02}&&\ding{51}&\ding{51}&\ding{51}\\ \hline

\multirow{ 3}{*}{\textsf{for the KM}}&
\textsf{KM-01}&\ding{51}&\ding{51}&\ding{51}&\ding{51}\\\cline{2-6} 
&\textsf{KM-02}&\ding{51}&\ding{51}&\ding{51}&\ding{51}\\ \cline{2-6}
&\textsf{KM-03}&\ding{51}&\ding{51}&\ding{51}&\ding{51}\\ \cline{2-6}
\hline
\multirow{ 8}{*}{\textsf{NFA}}& 

\textsf{NF-01}&&\ding{51}&\ding{51}&\\ \cline{2-6}
&\textsf{NF-02}&&\ding{51}&\ding{51}&\\ \cline{2-6}
&\textsf{NF-03}&\ding{51}&&&\ding{51}\\ \cline{2-6}
& \textsf{NF-04}&\ding{51}&\ding{51}&\ding{51}&\ding{51}\\ \cline{2-6}
&\textsf{NF-05}&&\ding{51}&\ding{51}&\\ \cline{2-6}
&\textsf{NF-06}&\ding{51}&\ding{51}&&\\ \cline{2-6}
&\textsf{NF-07}&&\ding{51}&\ding{51}&\ding{51}\\ \cline{2-6}
&\textsf{NF-08}&&\ding{51}&\ding{51}&\\ 
\hline
\end{tabular}
\caption{Coverage of the phases of the fault life cycle by the requirements}
\label{tab:lifecycle2req}
\end{center}
\end{table}

Furthermore, requirements regarding the \emph{organizational model} of the \textit{ioFMA} must be considered:

\begin{itemize}
\item \textbf{OM-01}: The \textbf{\textit{inter-organizational service delivery models}} have to be supported.
\item \textbf{OM-02}: \textbf{\textit{Definition of roles and responsibilities}} according to the use cases described above.
\end{itemize}

However, also the following requirements regarding the \emph{communication model} of the \textit{ioFMA} must be kept in mind:

\begin{itemize}
\item \textbf{KM-01}: \textbf{\textit{Communication mechanism}}, such as pull or push models have to be supported by the ioFMA.
\item \textbf{KM-02}: \textbf{\textit{Inter-domain communication}} is a very important requirement as the ioFMA will be deployed in an inter-organizational environment. Different networks with heterogeneous technologies exchange different data with each other. The inter-domain communication is also important, because in the absence of a central unit for coordination and communication between different networks at least a minimal set of information has to be exchanged.
\item \textbf{KM-03:} In order to support the data exchange within different networks a \textbf{\textit{communication protocol}} has to be defined. The complexity of the inter-organizational environment with their different provider, networks, and protocols is the challenge we are facing here.
\end{itemize}

Finally, we argue that the functional requirements regarding the sub-models of the \emph{ioFMA} must be complemented by the following series of non-functional requirements:

\begin{itemize}
\item \textbf{NF-01}: An \textbf{\textit{access control}} mechanism has to be part of the \emph{ioFMA}.
\item \textbf{NF-02}: Protection against data loss and deliberate data altering especially in the fault localization, reporting, and false-positive \textbf{\textit{data integrity}} has to be provided all the time.
\item \textbf{NF-03}: The \textbf{\textit{up-to-dateness}} of the data in the \emph{ioFMA} has to be guaranteed.
\item \textbf{NF-04}: Especially fault localization, monitoring, and false-positives management require a well-designed \textbf{\textit{scalability}} of the tools in order to provide the discussed functionality.
\item \textbf{NF-05}: Adequate \textbf{\textit{performance}} in the realization of the above named functionalities has to be achieved.
\item \textbf{NF-06}: The \textbf{\textit{automation}} of as many possible functionalities as possible has to be realized in order to speed up the fault resolution process.
\item \textbf{NF-07}: A \textbf{\textit{common data base}} for all the providers involved in the inter-organizational fault management process.
\item \textbf{NF-08}: Last but not least all processes and functionalities have to be properly \textbf{\textit{documented}}.
\end{itemize}

As we take the whole fault resolution process into account, the requirements have to be related to all relevant life cycle phases. Table \ref{tab:lifecycle2req} shows which requirements have to be fulfilled in the different phases of the fault life cycle (detection, isolation, repairing/recovery, and forecast/prevention).

\section{Core aspects of the functional model}\label{sec:6funcModel}

This section addresses the functional model of the ioFMA. As a base for its design the use cases described in section \ref{subsec:usecase} are applied. As stated in \cite{han99}, the functional model contains the functional areas which integrate all the required functionalities of a management architecture. For the ioFMA, we elicited three functional areas related to the organizational domain in which it is deployed: 

\begin{itemize}
 \item Provider management -- this the part of the ioFMA concerned with local ,,arrangements'' and integrating them with intra-organizational fault management
 \item Inter-organizational Management -- this is the core part of the functional model of the ioFMA as it contains all inter-organizational aspects 
 \item Customer management -- is placed on a more abstract level above the both former functional areas as it is connected to both of them and is the enabler of the provider and inter-organizational management, respectively. 
\end{itemize}

\subsection{Provider Management}

Within the service provider domain, different management functions in order to support the inter-organizational fault management have to be implemented. These management functions rely on the described use cases.

Fault localization within one's own domain is the first management function which has to be realized in a domain as a part of an ioFMA. The progress management for the fault resolution as well as the progress management for the maintenance work have to be performed within the service provider domain and connect to the inter-organizational management. Finding and removing false positives as well as performing data changes (under the strict control of the inter-organizational management) have also to be implemented within the domain.

\subsection{Inter-organizational Management}

As the core of the functional model, the inter-organizational management has to coordinate, integrate, put together information and functions from the different involved service provider domains. 
The management functions, which the inter-organizational management comprises, are: fault localization in an unspecified domain and within a specific domain, progress management for the fault resolution and for the maintenance work, overall monitoring and service monitoring, creation of statistical plots and accounting data reports, representation of QoS parameter and realization of trend analysis as well as detecting respective removing false positive fault reports. It can be observed that these are mainly the use cases defined previously. In addition to this a very important management function -- \emph{data change} -- has to be added. This has to be allowed but only under control of the inter-organizational management.

\subsection{Customer Management}

The customer management is the key enabler for both the provider management and the inter-organizational management. It actually contains all the management functions listed above, but has additional functionality. For example, from the customer's perspective the opening and updating of fault reports has to be supported. It serves as both a trigger and a feedback channel and is an essential core component of IT service management architectures.

\begin{figure}[] 
	\centerline{\includegraphics[width=0.95\columnwidth]{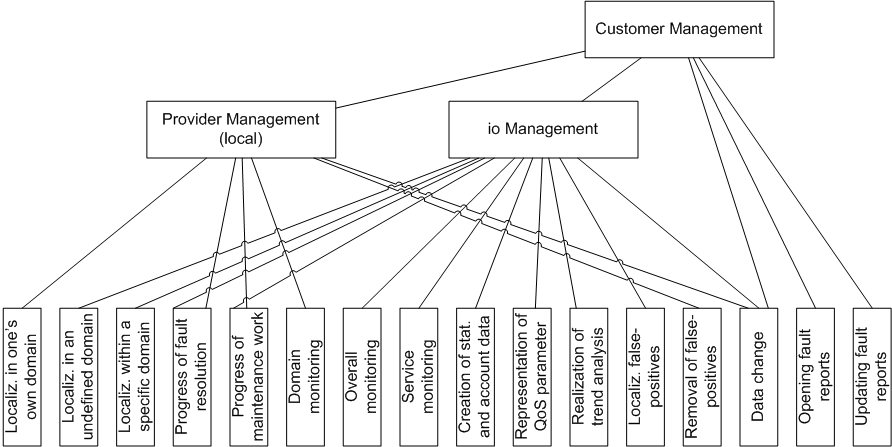}}
	\caption{Overview on the functional model of the ioFMA}
	\label{fig:functionalModel} 
\end{figure}

\section{Conclusions and future work}\label{sec:5further}

In this article we presented a full requirement analysis in order to design an inter-organizational fault management architecture. We also discussed the core aspects of the functional and organizational models based on the elicited use cases and requirements. The next steps in our research are to complete the architecture with a communication and an information model. After that we will deliver a full model of ioFMA on the PIM layer as well as its transformation to the system layer. Our implementation will be customized for the LHC optical private network (LHCOPN), which is operated by the European G\'EANT network.

\small

\section*{Acknowledgments}
The authors would like to thank their colleagues at the Leibniz Supercomputing Centre of the Bavarian Academy of Sciences and Humanities (see \url{http://www.lrz.de/}) for helpful discussions and valuable comments about this paper.

The authors wish to thank the members of the Munich Network Management Team (MNM-Team) for helpful discussions and
valuable comments on previous versions of this paper. The MNM Team directed by Prof. Dr. Dieter Kranzlm\"uller and Prof.
Dr. Heinz-Gerd Hegering is a group of researchers at Ludwig-Maximilians-Universit\"at M\"unchen, Technische Universit\"at
M\"unchen, the University of the Federal Armed Forces and the Leibniz Supercomputing Centre of the Bavarian Academy of
Sciences and Humanities. See \url{http://www.mnm-team.org/}.

\bibliographystyle{IEEEtran}
\bibliography{promobib}

\begin{thebibliography}{10}
\providecommand{\url}[1]{#1}
\csname url@samestyle\endcsname
\providecommand{\newblock}{\relax}
\providecommand{\bibinfo}[2]{#2}
\providecommand{\BIBentrySTDinterwordspacing}{\spaceskip=0pt\relax}
\providecommand{\BIBentryALTinterwordstretchfactor}{4}
\providecommand{\BIBentryALTinterwordspacing}{\spaceskip=\fontdimen2\font plus
\BIBentryALTinterwordstretchfactor\fontdimen3\font minus
  \fontdimen4\font\relax}
\providecommand{\BIBforeignlanguage}[2]{{%
\expandafter\ifx\csname l@#1\endcsname\relax
\typeout{** WARNING: IEEEtran.bst: No hyphenation pattern has been}%
\typeout{** loaded for the language `#1'. Using the pattern for}%
\typeout{** the default language instead.}%
\else
\language=\csname l@#1\endcsname
\fi
#2}}
\providecommand{\BIBdecl}{\relax}
\BIBdecl

\bibitem{itil07d}
{OGC}, Ed., \emph{Service Operation}, ser. IT Infrastructure Library v3 (ITIL
  v3).\hskip 1em plus 0.5em minus 0.4em\relax Norwich, UK: The Stationary
  Office, 2007.

\bibitem{etom}
``{enhanced Telecom Operations Map (eTOM), The Business Process Framework for
  the Information and Communications Services Industry},'' {TeleManagement
  Forum}, GB 921 Release 5.0, Apr. 2005.

\bibitem{Hamm09}
M.~Hamm, ``{IT Service Management Prozesse verketteter Dienste},''
  Dissertation, {Ludwig--Maximilians--Universit\"at M\"unchen}, Jun. 2009.

\bibitem{hmy08}
M.~Hamm, P.~Marcu, and M.~Yampolskiy, ``{Beyond Hierarchy: Towards a Service
  Model supporting new Sourcing Strategies for IT Services},'' in
  \emph{Proceedings of the 2008 Workshop of HP Software University Association
  (HP-SUA), Infonomics-Consulting, Hewlett-Packard}, Marrakech, Morocco, June
  2008.

\bibitem{han99}
H.-G. Hegering, S.~Abeck, and B.~Neumair, \emph{{Integrated Management of
  Networked Systems - Concepts, Architectures and their Operational
  Application}}.\hskip 1em plus 0.5em minus 0.4em\relax Morgan Kaufmann
  Publishers, 1999.

\bibitem{iso20000}
``{ISO/IEC 20000-1:2005 - Information Technology - Service Management - Part 1:
  Specification},'' {International Organization for Standardization}, Tech.
  Rep., Dec. 2005.

\bibitem{Dreo02}
G.~{Dreo Rodosek}, ``{A Framework for IT Service Management},'' Habilitation,
  University of Munich (LMU), Department of Computer Science, Munich, Germany,
  Jun. 2002.

\bibitem{Hed05}
G.~Hedlund, ``Assumptions of hierarchy and heterarchy, with applications to the
  management of the multinational corporation,'' in \emph{Organizational Theory
  and the Multinational Corporation}, 2nd~ed., S.~Ghoshal and E.~Westney, Eds.,
  London, 2005, pp. 198--221.

\bibitem{vpmb07}
R.~L. Vianna, E.~R. Polina, C.~C. Marquezan, L.~Bertholdo, L.~M.~R. Tarouco,
  M.~J.~B. Almeida, and L.~Z. Granville, ``{An Evaluation of Service
  Composition Technologies Applied to Network Management},'' in \emph{{10th
  IFIP/IEEE International Symposium on Integrated Network Management}},
  {Munich}, 2007, pp. 420--428.

\bibitem{kgf07}
T.~Klie, F.~Gebhard, and S.~Fischer, ``{Towards Automatic Composition of
  Network Management Web Services},'' in \emph{{Integrated Network Management,
  IM 2007. 10th IFIP/IEEE International Symposium on Integrated Network
  Management}}, {Munich, Germany}, 2007, pp. 769--772.

\bibitem{mdkp09}
S.~Mitra, P.~Dutta, S.~Kalyanaraman, and P.~Pradhan, ``{Spatio-Temporal
  Patterns for Problem Determination in IT Services},'' pp. 49--56, Sep. 2009.

\bibitem{GPM08a}
R.~Gupta, K.~H. Prasad, and M.~K. Mohania, ``Information integration techniques
  to automate incident management,'' in \emph{Proceedings of the IEEE/IFIP
  Network Operations and Management Symposium: Pervasive Management for
  Ubioquitous Networks and Services (NOMS 2008)}.\hskip 1em plus 0.5em minus
  0.4em\relax Salvador Bahia, Brazil: IFIP/IEEE, Apr. 2008, pp. 979--982.

\bibitem{MSGL09}
P.~Marcu, L.~Shwartz, G.~Grabarnik, and D.~Loewenstern, ``{Managing Faults in
  the Service Delivery Process of Service Provider Coalitions},'' in
  \emph{{IEEE International Conference on Service Computing (SCC 2009)}},
  Bangalore, India, Sep. 2009.

\bibitem{OMG03}
``{MDA Guide},'' http://www.omg.org/mda/, Jun 2003.

\bibitem{IntegraTUM}
``{IntegraTUM project, Technische Universit\"at M\"unchen},''
  http://portal.\-mytum.de/iuk/integratum/index\_html.

\bibitem{hokn10}
W.~Hommel and S.~Knittl, ``{Aufbau von organisations\"ubergreifenden
  Fehlermanagementprozessen im Projekt IntegraTUM},'' in
  \emph{Informationsmanagement in Hochschulen}, A.~Bode and R.~Borgeest,
  Eds.\hskip 1em plus 0.5em minus 0.4em\relax Berlin: Springer-Verlag, 2010.

\bibitem{Geant10}
{G\'{E}ANT}, ``{G\'{e}ANT Homepage},'' \url{ http://www.geant.net/}, 2010.

\bibitem{IPPM}
\BIBentryALTinterwordspacing
``{IP Performance Metrics Working Group}.'' [Online]. Available:
  \url{http://tools.ietf.org/wg/ippm/}
\BIBentrySTDinterwordspacing

\bibitem{RFC2679}
G.~Almes, S.~Kalidindi, and M.~Zekauskas, ``{A One-way Delay Metric for
  IPPM},'' {USA}, Tech. Rep., 1999.

\bibitem{RFC3393}
C.~Demichelis and P.~Chimento, ``{IP Packet Delay Variation Metric for IP
  Performance Metrics (IPPM)},'' {USA}, Tech. Rep., 2002.

\bibitem{RFC2680}
G.~Almes, S.~Kalidindi, and M.~Zekauskas, ``{A One-way Packet Loss Metric for
  IPPM},'' {USA}, Tech. Rep., 1999.

\end{thebibliography}

\subsection*{Autors}
\label{sec:shortBio}
\normalsize
\label{sec:PatriciaMarcu}
\parpic[r]{\includegraphics[width=0.2\textwidth]{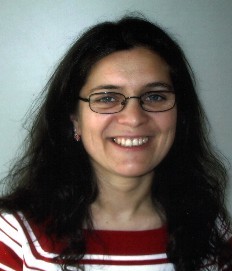}}
\textbf{Patricia Marcu} received her diploma in Computer Science in 2006 at the LMU Munich. In 2007 she joined the MNM-Team at Leibniz Supercomputing Centre as a research assistant and pursues her Ph.D. degree in Computer Science. She is currently working on the further development of the Customer Network Managemnt (CNM) tool and on the visualization of the LHCOPN within the European Geant project. Her research focuses on inter-organizational fault management and IT Service Management.
\\

\label{sec:WolfgangHommel}
\parpic[r]{\includegraphics[width=0.2\textwidth]{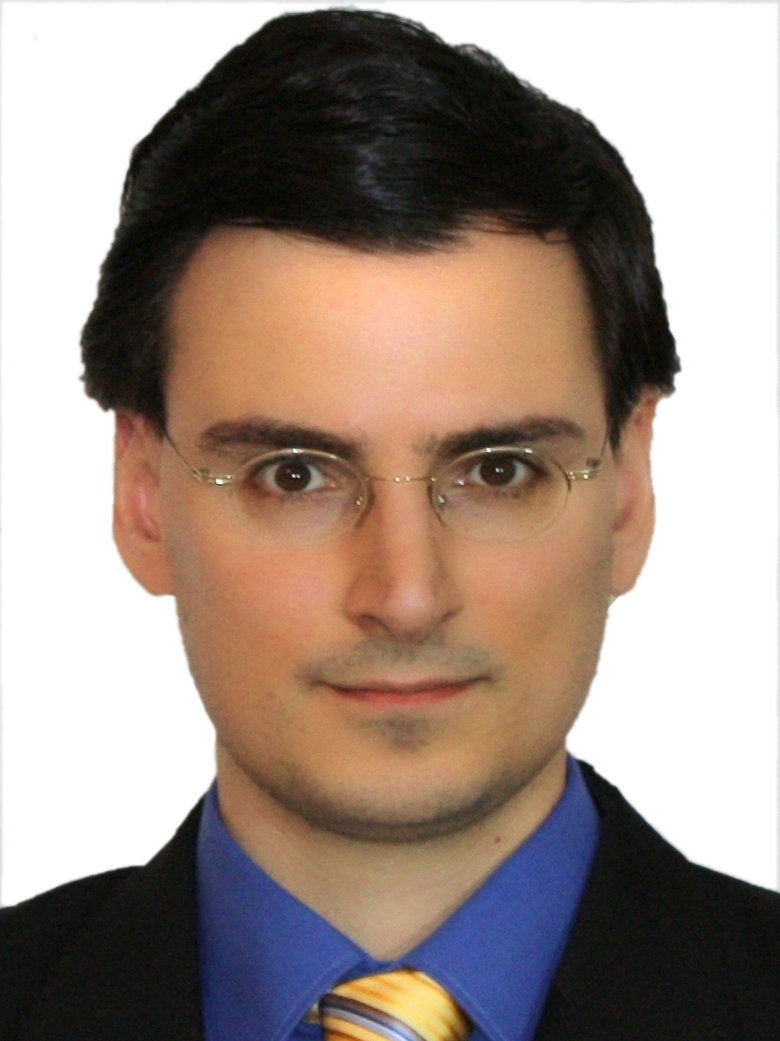}}

\textbf{Wolfgang Hommel} has a Ph.D. in computer science from LMU Munich, and heads the network services planning group at the Leibniz Supercomputing Centre. His current research focuses on IT security and privacy management in large distributed systems, including identity federations and Grids. Emphasis is put on a holistic perspective, i.e., the problems and solutions are analyzed from the design phase through software engineering, deployment in heterogeneous infrastructures, and during the
operation and change phases according to IT service management process frameworks, such as ISO/IEC 20000-1. 

\end{document}